\documentclass[nofootinbib, superscriptaddress, twocolumn]{revtex4-2}
\usepackage[utf8]{inputenc}
\usepackage{bm, amsthm, amsmath, amsfonts, amssymb, color, graphicx, natbib}
\usepackage{slashed}
\usepackage{xcolor}
\usepackage{verbatim}
\usepackage{float}
\graphicspath{{fig/}}
\usepackage{hyperref}
\hypersetup{
	colorlinks=true,
	linkcolor=blue,
	filecolor=blue,      
	urlcolor=blue,
	citecolor=blue
}

\begin{document}
	
	\title{Toward Cosmological Standard Timers in Primordial Black Hole Binaries}
	
	\author{Qianhang Ding}
	\email{qdingab@connect.ust.hk}
	\affiliation{Department of Physics, The Hong Kong University of Science and Technology, Hong Kong, P.R.China}
	\affiliation{Jockey Club Institute for Advanced Study, The Hong Kong University of Science and Technology, Hong Kong, P.R.China}

\begin{abstract}
	We propose that primordial black hole (PBH) binary systems can lead to standard timers in tracking the evolution of the Universe. Through gravitational waves from monochromatic PBH binaries, the probability distribution on major axis and eccentricity from the same redshift is obtained. By studying the dynamical evolution of PBH binaries from the initial probability distribution to observed redshifted ones, the redshift-time calibration can be extracted, which can constrain cosmological models. A general formalism of the standard timer is further concluded based on the evolution of statistical distribution in dynamical systems.
\end{abstract}

\maketitle

\section{Introduction}

With the development of modern cosmology, various cosmic properties have been found in observations, e.g., the cosmic microwave background (CMB) \cite{PhysRev.74.505.2, PhysRev.74.1737, PhysRev.75.1089, Planck:2018vyg}, large scale structure (LSS) \cite{Peebles:1982ff}, and cosmic accelerating expansion \cite{SupernovaCosmologyProject:1998vns, SupernovaSearchTeam:1998fmf}. Lambda-cold dark matter ($\Lambda$CDM) model is therefore established \cite{Rubin:1970zza, Faber:1979pp, PhysRevLett.52.2090, Frenk1985} as the standard model in cosmology.

For precise understanding in evolution of the Universe, various observational technologies have been proposed in studying the cosmological distance-redshift relation. Type Ia supernovae (SNe) produce consistent peak luminosity. The luminosity distance can be obtained by comparing their absolute and apparent magnitude, which makes sure Type Ia SNe work as the standard candle to provide luminosity distance-redshift relation \cite{Fernie_1969, Pan-STARRS1:2017jku}. Baryon acoustic oscillations (BAO) determine a fixed sound horizon, by observing sound horizons at different redshifts. The calibration between angular diameter distance and redshift is extracted, which serves as standard rulers \cite{beutler20116df, Ross:2014qpa, BOSS:2014hwf, Planck:2018vyg}. Gravitational waves (GWs) from binary systems and their electromagnetic counterparts provide the calibration between the luminosity distance and redshift as standard sirens \cite{LIGOScientific:2017vwq}. However, there still exist puzzles in $\Lambda$CDM model, e.g., the Hubble tension \cite{Riess:2021jrx, Riess:2016jrr, Planck:2016kqe} and the $S_8$ tension \cite{Planck:2018vyg, Nunes:2021ipq}. Such tensions could be caused by unrecognized systematic uncertainties in measurements \cite{Freedman:2017yms, Rameez:2019wdt} or hidden new physics \cite{Poulin:2018cxd, Bernal:2016gxb, Lin:2019qug, Kenworthy:2019qwq, Ding:2019mmw, Vagnozzi:2019ezj, Fung:2021fcj}. New observational methods are needed in cross checking the observed tensions.

Intuitively, the evolution of the Universe can be briefly characterized by the scale factor $a(t)$ in the Friedmann-Robertson-Walker (FRW) metric, which is related to the cosmological redshift $z$ by $1 + z(t) = a_0/a(t)$, where $a_0$ is the present scale factor. Tracking $z(t)$ provides another perspective in studying cosmological evolution (also used in the determination of the age of the Universe, see \cite{Janes:1983jx, Chaboyer:1995tw, LB2003, Boylan-Kolchin:2021fvy} for details), which could be achieved in the cosmological dynamical systems. Following their intrinsic dynamics, the physical evolution time of dynamical systems from the initial state to the later state is attained. Meanwhile, redshift is decoded from their observable. Due to the independence of physical evolution time of dynamical systems in the evolution of the Universe, it works as a timer in calibrating the redshift-time relation. This approach is known as {\it standard timers}.

In the first study on standard timers \cite{Cai:2021fgm}, we have shown that through the Hawking radiation emitted from light primordial black hole (PBH) clusters, PBH stellar bubbles \cite{Cai:2021zxo} can be a potential candidate of standard timers. Due to the primordial origin of PBHs, the initial mass function of PBHs in clustering should be the same. With the emission of Hawking radiation, PBHs evaporate which deforms the mass function of PBHs. By studying the evolution of PBH mass function, its physical evolution time is attained. Meanwhile observing the gamma ray spectrum from PBH stellar bubbles gives the redshifted PBH mass function where redshift is encoded. Hence, the calibration between redshift and physical evolution time of PBH mass function is constructed.

In this paper, we note that PBH binary systems can lead us to standard timers. Under the assumption of random distribution of PBHs in space \cite{Ballesteros:2018swv, MoradinezhadDizgah:2019wjf, Inman:2019wvr},  the PBH binaries could decouple from the Hubble flow and form an identical initial probability distribution on the major axis $a$ and the eccentricity $e$ \cite{Sasaki:2016jop, Ali-Haimoud:2017rtz} (see Eq.~\eqref{eq:PBH_initial_dist} for example). With the emission of GWs, the major axis and eccentricity shrink, which changes the later probability distribution in PBH binary systems accordingly. Through the waveform of GWs from the same redshift PBH binaries, the probability distribution on major axis and eccentricity from the same redshift is obtained. By studying the evolution of PBH binaries from the initial probability distribution to later ones, cosmological redshift and physical evolution time in PBH binary systems are connected, which can constrain cosmological models. 

Although PBHs are hypothetical objects, there still exist some GW events which may indicate the existence of PBHs. For instance, GW190521 shows the mass of binary components lies in the astrophysical BH mass gap \cite{LIGOScientific:2020iuh} and such a event could be explained in PBHs scenario \cite{DeLuca:2020sae}, GW190425 and GW190814 show a companion of BH binary has a mass smaller than $3 M_\odot$ \cite{LIGOScientific:2020aai, LIGOScientific:2020zkf} and such events could origin from PBHs scenario \cite{Clesse:2020ghq}. Considering the next generation of GW detectors, such as the Einstein Telescope \cite{Punturo:2010zz}, Laser Interferometer Space Antenna \cite{bender1998lisa}, can detect GWs from high redshift ($z > 20$), such a high redshift binary BHs event could be a smoking gun for the existence of PBHs \cite{Nakamura:2016hna, Ding:2020ykt, Ng:2022agi}. Also, PBH binaries may dominate the high redshift binary systems, detecting GWs from PBH binary systems at high redshift would own a pure GW background, which can help construct the high-precision standard timer.

This paper is organized as follows. In Sec.~\ref{sec:standard_timer_PBH_binary}, we show how to construct the standard timer in PBH binary systems, including single parameter PBH binary systems in Sec.~\ref{sec:single_parameter}, multi-parameter PBH binary systems in Sec.~\ref{sec:multi_parameter} and PBH binary systems without the initial probability distribution in Sec.~\ref{sec:standard_timer_no_initial_state}. In Sec.~\ref{sec:conclusion_discussions}, the conclusion and discussions about the standard timer from PBH binary systems are given. In Appendix \ref{appendix:single_parameter} \& \ref{appendix:multi_parameter}, a formalism on constructing standard timers in general dynamical systems is shown.

\section{Toward standard timers in PBH binary systems}\label{sec:standard_timer_PBH_binary}

In constructing the standard timer in PBH binary systems, two essential requirements are needed. One is the identical initial state of PBH binary systems, which works as a standard reference in extracting physical evolution time.
\footnote{The identity of PBH initial state is a key requirement, which makes sure the physical evolution time in PBH binary systems can be an independent measurement for cosmic time.}
The other one is that their later evolved state from the same redshift can be obtained, which makes sure the redshift can be decoded from their observable. The former of requirements is achieved in the identical initial probability distribution on major axis and eccentricity  \cite{Sasaki:2016jop, Ali-Haimoud:2017rtz}
under the assumption of random distribution of PBHs in the space. The latter can be easily realized in local signal sources, such as the PBH cluster \cite{Belotsky:2018wph, Ding:2019tjk, Cai:2021zxo}, where the redshift of signals from a local source should be the same. However, redshifts of PBH binaries are hardly classified, due to their being globally distributed in the Universe.

The potential monochromatic mass spectrum in PBH scenarios changes the story. It introduces PBH binary systems in the standard timer through GW channels. After collecting many GW signals from different redshifts, we can extract the redshifted chirp mass $\mathcal{M}_z$ and mass ratio $q$ from the GW waveform. Due to the unknown intrinsic chirp mass $\mathcal{M}$, which follows $\mathcal{M} = \mathcal{M}_z/(1 + z)$ \cite{Finn:1992xs}, the redshift of binary systems cannot be determined. However, if GW signals come from PBH binaries, under the assumption that the mass of PBHs is monochromatic, the mass ratio of PBH binaries follows $q = 1$ (also a number of BH binaries with $q = 1$ are found in GWTC-2, see \cite{LIGOScientific:2020ibl} for details), and GW signals emitted from PBH binaries at the same redshift can give the same redshifted chirp mass. Then, PBH binaries can be classified into different redshift shells based on their redshifted chirp mass. As the result, the standard timer can be constructed by comparing the initial probability distribution on major axis $a$ and eccentricity $e$ and later ones from the same redshift.

\subsection{A toy model in single parameter PBH binary systems}\label{sec:single_parameter}

By assuming the mass of PBHs is monochromatic, the state of PBH binary systems can be described by a probability distribution on major axis $a$ and eccentricity $e$, which is $dP/dade$. For an intuitive understanding on how standard timers work in PBH binary systems, we start with circular binary systems (some discussions propose the circularising of PBH binaries before entering GW frequency bands, see \cite{Franciolini:2021xbq} for details). Then states of PBH binary systems only depend on the single parameter major axis $a$, which is $dP/da$. In studying the evolution of its probability distribution, we have
\begin{align}
	S(a; t) = \frac{dP}{da_t} = \frac{dP}{da_\mathrm{i}}\frac{da_\mathrm{i}}{da_t} = S(a; t_\mathrm{i})\frac{da_\mathrm{i}}{da_t}~,
\end{align}
where,  $a_\mathrm{i}$ and $a_t$ denote the major axis at the initial physical time $t_\mathrm{i}$ and later physical time $t$, respectively. $S(a; t)$ denotes the probability distribution on major axis $a$ at physical time $t$ and $dP/da_\mathrm{i}$ is the initial probability distribution $S(a; t_\mathrm{i})$. In connecting $S(a; t)$ with $S(a; t_\mathrm{i})$, we consider time evolution of the major axis, following \cite{Peters:1964zz}
\begin{align}\label{eq:axis_time_evolution}
	\frac{da}{dt} = -\frac{128}{5}\frac{G^3 M_\mathrm{PBH}^3}{c^5 a^3}~,
\end{align}
where $M_\mathrm{PBH}$ is the monochromatic mass of PBHs, $G$ and $c$ are the Newton's constant and the speed of light, respectively. Integrating Eq.~\eqref{eq:axis_time_evolution} from the initial physical time $t_\mathrm{i}$ to later physical time $t$, we have the evolution of the major axis,
\begin{align}\label{eq:major_axis_fun}
	a_\mathrm{i}^4 = a_t^4 + \delta^4(\Delta t)~,
\end{align}
where $\delta^4$ is defined as $\delta^4(\Delta t) \equiv 512 G^3 M_\mathrm{PBH}^3 \Delta t/5 c^5$ and $\Delta t$ is the physical evolution time $\Delta t = t - t_\mathrm{i}$. Because the initial probability distribution $S(a; t_\mathrm{i})$ is fixed, it makes sure the physical evolution time $\Delta t$ can be obtained from the later probability distribution $S(a; t)$ and their internal dynamic. Due to the independence of the physical evolution time in the evolution of the Universe, $\Delta t$ can work as an independent cosmic time measurement between cosmic time $t_\mathrm{i}$ and $t$. Then, we differentiate Eq.~\eqref{eq:major_axis_fun} on both sides, the evolution of major axis $da_\mathrm{i}/da_t$ can be obtained. The probability distribution of single parameter PBH binary systems from an identical redshift shell can be expressed as following
\begin{align}\label{eq:distribution_inter}
	S(a; t) = \frac{dP}{da_\mathrm{i}}\frac{a_t^3}{(a_t^4 + \delta^4(\Delta t))^{3/4}}~.
\end{align}

With the expansion of the Universe, the observed major axis $a_z$ is redshifted by $a_z = (1 + z) a$, which can be found in the Kepler's third law $a_z \sim (G M_z)^{1/3} f_{z}^{-2/3} \sim (1 + z) (G M)^{1/3} f^{-2/3}$, where $M_z$ and $f_z$ are the observed mass and GW frequency in binary systems, respectively. Therefore, the observed probability distribution follows
\begin{align}
	S_o(a_z; t) = \frac{dP}{da_\mathrm{i}(z)}\frac{a_z^3}{(a_z^4 + \delta^4(\Delta t_z))^{3/4}}~.
\end{align}
Here, subscript $o$ denotes the observational quantity. As Eq.~\eqref{eq:major_axis_fun}, we have the relation $a_\mathrm{i}^4(z) = a_z^4 + \delta^4(\Delta t_z)$, where $\Delta t_z$ depends on the redshift of PBHs binaries. In order to obtain the physical evolution time $\Delta t$ from $\Delta t_z$, a correct redshift should be firstly obtained from the observed probability distribution. We consider the condition $a_z^4 \gg \delta^4(\Delta t_z)$, which infers $a_\mathrm{i}(z) \simeq a_z$. In this large major axis limit, the evolution of major axis is negligible. Therefore, the observed distribution of the major axis will change only due to the cosmic expansion. The observed probability distribution becomes
\begin{align}
	S_o(a_z; t) \simeq S_o(a_\mathrm{i} (1 + z); t) = \frac{dP}{da_\mathrm{i}(z)} = \frac{1}{1 + z}\frac{dP}{da_\mathrm{i}}~.
\end{align}
If we consider a specific major axis $a_L$ in the limit of large major axis, we obtain a equation on redshift
\begin{align}\label{eq:solve_redshift}
	(1 + z) S_o(a_L(1 + z); t) = S(a_L; t_\mathrm{i})~.
\end{align}
Here, we have already known $S_o(a_z; t)$ in observations and the initial probability distribution $S(a; t_\mathrm{i})$, then redshift can be numerically solved in Eq.~\eqref{eq:solve_redshift}. After obtaining the redshift, the intrinsic probability distributions can recover from the redshifted ones by $S(a; t) = (1 + z)S_o(a(1 + z); t)$ and the mass of PBHs can be solved from the redshifted mass by $M_\mathrm{PBH} = M_z/(1 + z)$. Then, physical evolution time can be extracted in the condition $a_t^4 \ll \delta^4(\Delta t)$, which indicates $a_\mathrm{i} \simeq \delta(\Delta t)$. In this small major axis limit, we have the log probability distribution from Eq.~\eqref{eq:distribution_inter} as following
\begin{align}\label{eq:log_distribution}
	\log{S(a; t)} \simeq \log{\frac{S(\delta(\Delta t); t_\mathrm{i})}{\delta^3(\Delta t)}} + 3 \log{a_t}~.
\end{align}
Then $\delta(\Delta t)$ can be extracted from $S(\delta(\Delta t); t_\mathrm{i})/\delta^3(\Delta t)$ in Eq.~\eqref{eq:log_distribution}. $\delta(\Delta t)$ depends on the mass of PBHs $M_\mathrm{PBH}$ and physical evolution time $\Delta t$. In above calculation, the mass of PBHs has been found after obtaining redshift. Then, physical evolution time $\Delta t$ can be resolved from obtained $\delta(\Delta t)$. After obtaining the redshift and physical evolution time, the calibration between redshift and time can be constructed in single parameter PBH binary systems.

\begin{figure}\centering
	\includegraphics[width=8cm]{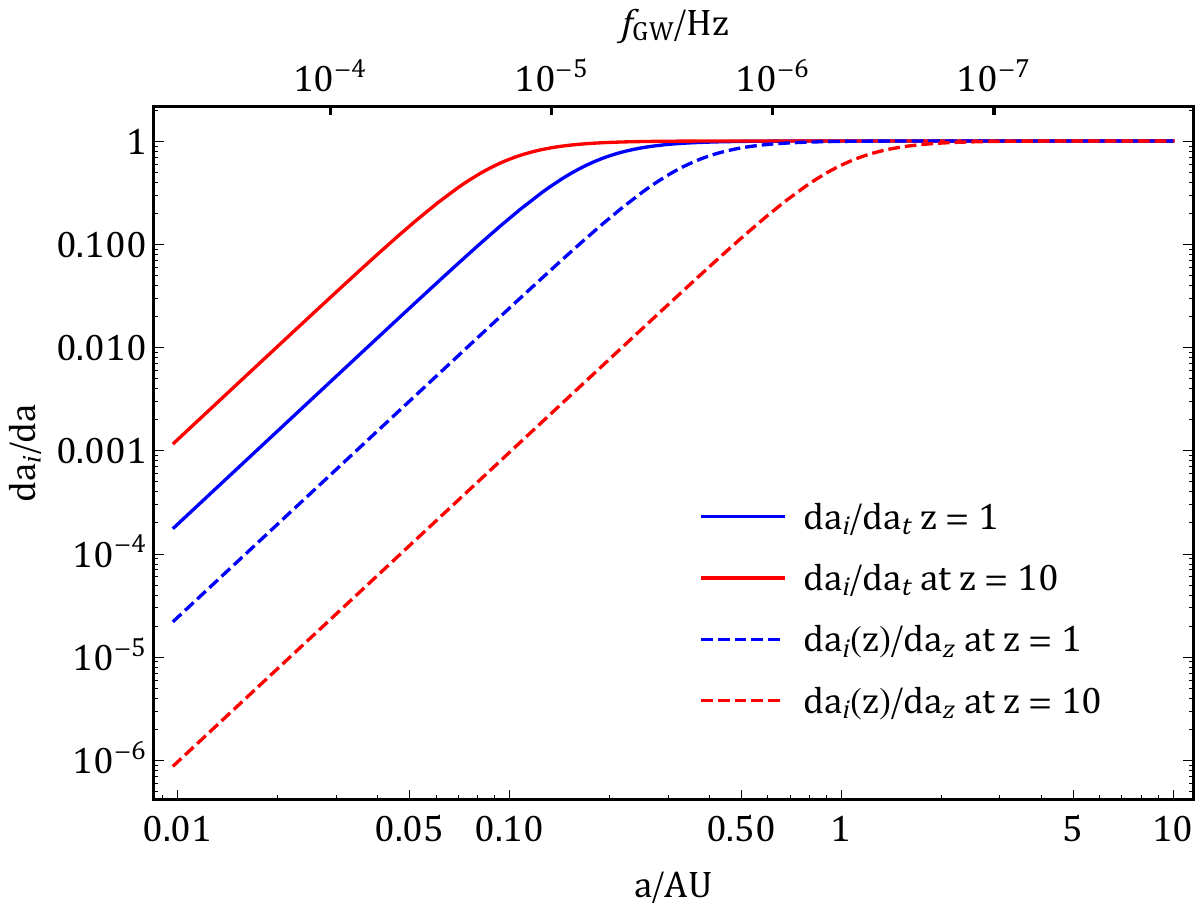}
	\caption{\label{fig:evolution_function}
		The evolution of single parameter PBH binary systems $da_\mathrm{i}/da$. We assume the mass of PBHs is $30 M_\odot$ and physical evolution time is the standard cosmological time from $z = 3000$ to the redshift denoted in the figure. The solid curve is the intrinsic evolution of dynamical systems and the dashed curve is the observed evolution of dynamical systems.
	}
\end{figure}

In general, the standard timer requires two properties in the evolution of dynamical systems, we take the single parameter PBH binary system as an example to show that in Fig.~\ref{fig:evolution_function}.
Typical properties in $da_\mathrm{i}/da$ are the flat constant part in the large major axis tail and the rapid evolution part in the small major axis tail. The flat constant part is the region where time evolution is negligible, the redshift can be extracted by comparing the redshifted and initial probability distributions in this region. After obtaining the redshift, intrinsic evolution functions (solid lines) can be recovered from redshifted ones (dashed lines) in Fig.~\ref{fig:evolution_function}. The rapid evolution part describes the physical time evolution effectively changing the major axis of PBH binaries, where physical evolution time can be extracted through Eq.~\eqref{eq:log_distribution}. In principle, redshift-time relation can be decoded from observed probability distributions, which indicates redshift and physical evolution time should be hidden in different parts of observed distribution, otherwise redshift-time degeneracy in observed distributions could cause the uncertainty in extracting redshift-time relation. Therefore, the statistical distribution of a standard timer candidate should include inactive evolution part for extracting redshift and rapid evolution part for extracting physical evolution time.

In addition, an applicable standard timer should constrain redshift-time relation to the precision higher than the result from the $\Lambda$CDM model, namely, $\mathcal{O}(0.1)$, due to the Hubble parameter difference between early measurements \cite{Planck:2018vyg} and late measurements \cite{Riess:2021jrx}. The measurement uncertainty of the standard timer is mainly from the uncertainty of redshifted BH mass, which is around $\mathcal{O}(0.1)$ (see TABLE VI of \cite{LIGOScientific:2020ibl}), apply this uncertainty in the Kepler's third law, Eq.~\eqref{eq:solve_redshift} and \eqref{eq:log_distribution}, we can obtain the uncertainty of redshift-time relation from the standard timer is comparable with the $\Lambda$CDM result. With the better sensitivity of GW detectors, more GW waveform templates, and detailed studies and observations on PBH properties in the future, this potential standard timer would be improved to have higher precision and put strong constraints on the cosmic evolution.
  
\subsection{A practical model in PBH binary systems}\label{sec:multi_parameter}

A practical description of monochromatic PBH binary systems needs two parameters, major axis $a$ and eccentricity $e$.
In multi-parameter probability distributions, the evolution of probability distribution of PBH binary systems $dP/dade$ from an identical redshift shell can be described as following 
\begin{align}\label{eq:Jacobian}\nonumber
	&S(a, e; t) = \frac{dP}{da_\mathrm{i} de_\mathrm{i}} \det{\mathbf{J}(a, e, \Delta t)}~,\\
	&\mathbf{J}(a, e, \Delta t) = 
	\begin{pmatrix}
		\frac{\partial a_\mathrm{i}}{\partial a_t} & \frac{\partial a_\mathrm{i}}{\partial e_t}\\
		\frac{\partial e_\mathrm{i}}{\partial a_t} & \frac{\partial e_\mathrm{i}}{\partial e_t}\\
	\end{pmatrix}~.
\end{align}
Here, $dP/da_\mathrm{i} de_\mathrm{i}$ is the initial probability distribution of PBH binary systems, and $\mathbf{J}(a, e, \Delta t)$ is the Jacobian of two-parameter PBH binary systems after the evolution of physical evolution time $\Delta t$, which connects the initial and later probability distributions. In calculating $\mathbf{J}(a, e, \Delta t)$, we consider the time evolution of parameters in PBH binaries, following \cite{Peters:1964zz}
\begin{align}\label{eq:a_e_evolution}\nonumber
	\frac{da}{dt} &= -\frac{128}{5}\frac{G^3 M_\mathrm{PBH}^3}{c^5 a^3 (1 - e^2)^{7/2}}\left(1 + \frac{73}{24}e^2 + \frac{37}{96}e^4\right)~,\\
	\frac{de}{dt} &= -\frac{608}{15}\frac{G^3 M_\mathrm{PBH}^3}{c^5 a^4} \frac{e}{(1 - e^2)^{5/2}} \left(1 + \frac{121}{304}e^2\right)~.
\end{align}

Due to the expansion of the Universe, the cosmological redshift is introduced in the observed probability distribution, which is
\begin{align}\label{eq:multi_obs_dist}
	S_o(a_z, e; t) = \frac{dP}{da_\mathrm{i}(z) de_\mathrm{i}} \det{\mathbf{J}(a_z, e, \Delta t_z)}~,
\end{align}
where redshifted major axis becomes $a_z = (1 + z) a$. However, the redshift does not leave imprints on the eccentricity $e$, it can be found the $1+z$ factors coming from the major axis, mass and time in Eq.~\eqref{eq:a_e_evolution} cancel out with each other, which results in no redshift effect appearing in observed eccentricity. In measuring eccentricity in binary systems, the precision is not very high due to the lack of suitable GW waveform templates \cite{Gayathri:2020coq}. Therefore, we only consider probability distributions on the major axis $dP/da$ for a practical numerical solution, which can be obtained from Eq.~\eqref{eq:multi_obs_dist} as following
\begin{align}\label{eq:distribution_on_major_axis}
	S_o(a_z; t) = \frac{dP}{da_z} = \int_0^{e_\mathrm{max}} \frac{dP}{da_\mathrm{i}(z) de_\mathrm{i}} \det{\mathbf{J}(a_z, e, \Delta t_z)} de~.
\end{align}

\begin{figure*}\centering
	\includegraphics[width=18cm]{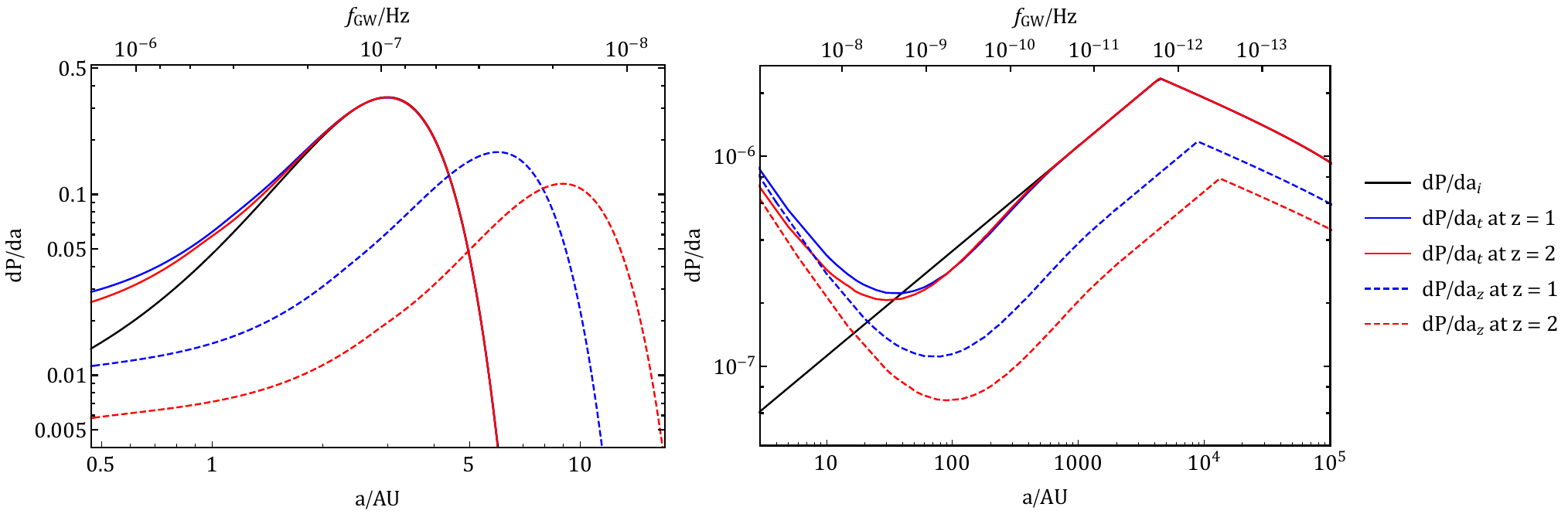}
	\caption{\label{fig:practical_model}
		The probability distribution of PBH binaries on the major axis $a$. We assume the mass of PBHs is $30 M_\odot$ and physical evolution time is the standard cosmological time from $z = 3000$ to the redshift denoted in the figure. Solid curves denote intrinsic probability distributions and dashed curves are observed probability distributions in PBH binary systems. {\it Left panel}: The initial probability distribution in PBH binary systems is the Gaussian distribution localized at $a = 3 \mathrm{AU}$ and $e = 0.9$, its standard deviations are $\sigma_a = 1 \mathrm{AU}$ and $\sigma_e = 0.05$. {\it Right panel}: The initial probability distribution follows Eq.~\eqref{eq:PBH_initial_dist}, where we set $f_\mathrm{PBH} = 0.1$ and $\bar{x} \simeq 1.3 \times 10^5 \mathrm{AU}$ in the $\Lambda$CDM model with $\Omega_\mathrm{DM} = 0.3$ and $H_0 = 73.2 \mathrm{km}\mathrm{s}^{-1}\mathrm{Mpc}^{-1}$. The maximal eccentricity $e_\mathrm{max}$ in Eq.~\eqref{eq:distribution_on_major_axis} is set as $\min{(0.9999, \sqrt{1 - (f_\mathrm{PBH} a/\bar{x})^{3/2}})}$, where upper bound $0.9999$ is for preventing the extreme nonlinear effect in probability distributions in numerical calculation.
	}
\end{figure*}

In the numerical solution of the evolution of probability distributions, we consider two types of initial probability distributions for the major axis and eccentricity of the PBH binaries. One is the Gaussian distribution localized at a specific major axis and eccentricity. After studying the Gaussian distribution, a general initial distribution can be decomposed into Gaussian distributions \cite{bhattacharya1967simple, gregor1969algorithm}. The other one is the standard probability distribution in PBH binary systems under the assumption of random spatial distribution of PBHs, which is chosen from \cite{Sasaki:2016jop, Ioka:1998nz},
\footnote{Even though initial probability distribution of PBH binaries is undetermined, the chosen distribution functions are reasonable in analysis. The analysis with initial Gaussian distribution can be extended to a general initial distribution case, and the analysis with the standard probability distribution under the random distribution of PBHs is an expected template in future GW data analysis.}
\begin{align}\label{eq:PBH_initial_dist}
	\frac{dP}{dade} = \frac{3}{4}f_\mathrm{PBH}^{3/2}\frac{a^{1/2}}{\bar{x}^{3/2}}\frac{e}{(1 - e^2)^{3/2}}~,
\end{align}
where $f_\mathrm{PBH}$ is the present energy density fraction of PBHs in the dark matter and $\bar{x}$ is the physical mean separation of PBHs at matter-radiation equality. Based on different types of initial probability distributions, a numerical study on the evolution of probability distribution of PBH binaries on the major axis $a$ following Eqs.~\eqref{eq:a_e_evolution}\eqref{eq:distribution_on_major_axis} is shown in Fig~.\ref{fig:practical_model}.

As mentioned in Sec.~\ref{sec:single_parameter}, various regions of probability distributions behave differently. In the large major axis limit, the evolution of probability distribution is negligible, the redshift can be obtained by comparing the initial probability distribution (black line) and redshifted probability distributions (dashed lines) in Fig~.\ref{fig:practical_model}, which can be further numerically resolved from the equation $(1 + z) S_o(a_L(1 + z); t) = S(a_L; t_\mathrm{i})$, where $a_L$ is picked from the large major axis region. However, the large major axis tail in observed probability distributions can hardly be obtained due to its extremely low gravitational wave frequency, as shown in the right panel of Fig.~\ref{fig:practical_model}. In this case, we can use numerical solutions as templates of probability distributions to match observational results. After obtaining the redshift, observed probability distributions $S_o(a_z; t)$ can be mapped to intrinsic probability distributions $S(a; t)$ following $S(a; t) = (1 + z) S_o(a(1 + z); t)$, and the mass of PBHs can be obtained from the redshifted mass following $M_\mathrm{PBH} = M_z/(1 + z)$. Then physical evolution time $\Delta t$ can be extracted from the numerical solution in the small major axis limit, where the evolution of probability distribution dominates. Accordingly, the redshift-time calibration is constructed in multi-parameter PBH binary systems. In addition, some extra effects could also influence the orbital evolution of PBH binaries, such as accretion of PBH binaries during the inspiral phase \cite{DeLuca:2020bjf, DeLuca:2020qqa}, the interaction between PBH binary with CDM particles \cite{Pilipenko:2022emp}. In building a realistic standard timer, we need to add extra contribution terms in Eq.~\eqref{eq:a_e_evolution} and numerically solve the Jacobian of orbital parameter evolution in Eq.~\eqref{eq:Jacobian}, then a high-precision and optimistic standard timer would be constructed.

Generally, the mass of PBHs is very essential in describing the evolution of PBH binary systems, when the mass function of PBHs is not monochromatic. The evolution of orbital parameters follows \cite{Peters:1964zz}
\begin{align}\label{eq:a_e_m_evolution}\nonumber
	\frac{da}{dt} &= -\frac{64}{5}\frac{G^3 m_1 m_2 (m_1 + m_2)}{c^5 a^3 (1 - e^2)^{7/2}}\left(1 + \frac{73}{24}e^2 + \frac{37}{96}e^4\right)~,\\
	\frac{de}{dt} &= -\frac{304}{15}\frac{G^3 m_1 m_2 (m_1 + m_2)}{c^5 a^4 (1 - e^2)^{5/2}} e \left(1 + \frac{121}{304}e^2\right)~.
\end{align}
Here, $m_1$ and $m_2$ are the mass of PBHs in the binary. Therefore the probability distribution of general PBH binary systems should be described as $dP/da de dM_\mathrm{G}$, where $M_\mathrm{G} \equiv (m_1 m_2 (m_1 + m_2))^{1/3}$. The evolution of probability distribution can also be numerically studied as we have discussed above. However, considering an extended mass function in PBH binary systems, it is hard to determine whether collected GWs come from the same redshift or not, because the redshifted mass depends on the mass of PBHs in binary and cosmological redshift, due to the mass uncertainty in extended mass spectrum of PBHs, the redshift is hardly determined. In inflationary scenario, the lognormal type mass function is a natural result of PBHs mass function \cite{Dolgov:1992pu, Carr:2016drx, Carr:2017jsz}, which is described by characteristic mass $M_c$ and width of mass spectrum $\sigma$. In the small $\sigma$ limit, the mass function of PBHs can be approximated by a monochromatic mass function, the redshift can be determined as we have discussed in the monochromatic case. Otherwise, the determination of redshift needs further studies.

\subsection{Toward standard timers without the initial probability distribution}\label{sec:standard_timer_no_initial_state}

In constructing standard timers in PBH binary systems, the initial probability distribution of PBH binaries plays an important role in extracting the redshift and physical evolution time by comparing it with observed redshifted probability distributions. However, the initial probability distribution on major axis and eccentricity of PBH binaries is indeterminate, due to the unknown mass spectrum and space distributions of PBHs, etc. Some previous works \cite{Ioka:1998nz, Ali-Haimoud:2017rtz, Chen:2018czv} have been done in calculating the initial probability distribution in different scenarios.

Therefore, toward standard timers without the initial probability distribution should be considered. Its feasibility is ensured that the initial probability distribution of PBH binaries is fixed after binaries form, under the assumption of random distribution of PBHs in space. It makes sure one-to-one correspondence between later evolved probability distribution and physical evolution time. We first consider circular PBH binary systems for an intuitive understanding. Due to lack of the initial probability distribution, we consider the PBH binaries from two different redshifts, their observed probability distributions are $S_o(a_{z_1}, t_1)$ and $S_o(a_{z_2}, t_2)$, respectively. By studying their dynamics, the connection between $z_1$, $z_2$ and $\Delta t = t_2 - t_1$ could be obtained.

In the large major axis limit, the evolution of major axis is negligible, we have the relation $S(a; t_\mathrm{i}) = (1 + z_1) S_o(a (1 + z_1); t_1) = (1 + z_2) S_o(a (1 + z_2); t_2)$, which gives the following equation,
\begin{align}\label{eq:redshift_ratio}
	S_o(a_L; t_1) = \eta S_o(\eta a_L; t_2)~,
\end{align}
where $a_L$ is picked up from the large major axis region, and $\eta$ is the redshift ratio which is defined as $\eta \equiv (1 + z_2)/(1 + z_1)$. With the observed $S_o(a_{z_1}; t_1)$ and $S_o(a_{z_2}; t_2)$, the redshift ratio can be numerically solved in Eq.~\eqref{eq:redshift_ratio}.

In the small major axis limit, the evolution of major axis is rapid. Following Eq.~\eqref{eq:major_axis_fun}, the relation between $a_{z_1}$ and $a_{z_2}$ is $a_{z_2}^4 = a_{z_1}^4 + \delta^4(\Delta t_{z_1})$. Then the relation between $S_o(a_{z_1}; t_1)$ and $S_o(a_{z_2}; t_2)$ can be expressed as
\begin{align}\label{eq:S12_relation}\nonumber
	S_o(a_{z_1}; t_1) &= \frac{dP}{da_{z_2}}\frac{da_{z_2}}{da_{z_1}}\\ &= S_o(a_{z_2}; t_2)\frac{a_{z_1}^3}{(a_{z_1}^4 + \delta^4(\Delta t_{z_1}))^{3/4}}~.
\end{align}
Here, $\Delta t_{z_1}$ depends on unknown redshift $z_1$. At the small major axis limit, $a_{z_1}^4 \ll \delta^4(\Delta t_{z_1})$, Eq.~\eqref{eq:S12_relation} can be approximated as following
\begin{align}\label{eq:deltatz}
	S_o(a_{z_1}; t_1) \simeq  \frac{S_o(\delta(\Delta t_{z_1}); t_2)}{\delta^3(\Delta t_{z_1})}a_{z_1}^3~.
\end{align}
In Eq.~\eqref{eq:deltatz}, $\delta(\Delta t_{z_1})$ can be numerically solved. From the observed $S_o(a_{z_1}; t_1)$ and $S_o(a_{z_2}; t_2)$, we have obtained the redshift ratio $\eta$ and $\delta(\Delta t_{z_1})$. In order to construct the redshift-time calibration $(z_1, z_2, \Delta t)$, the redshift $z_1$ and $z_2$ need to be determined. We can assume the cosmological evolution between two redshifts $z_1$ and $z_2$ follows the standard $\Lambda$CDM cosmology. Then, we assume that the redshift of one observed probability distribution is $\tilde{z}_1$ and the redshift of the other one is $\tilde{z}_2 = (1 + \tilde{z}_1) \eta - 1$. Under the assumption of $\tilde{z}_1$ and $\tilde{z}_2$, the PBH mass $M_\mathrm{PBH}$ can be first determined by $M_\mathrm{PBH} = M_{z_1}/(1 + \tilde{z}_1)$, and then physical evolution time $\Delta t$ can be obtained from $\delta(\Delta t_{z_1})$. The cosmological time between $\tilde{z}_1$ and $\tilde{z}_2$ can be calculated as $	t_z = \int_{\tilde{z}_1}^{\tilde{z}_2} dz/H(z)(1 + z)$, where $H(z)$ is the Hubble parameter along the line of sight. Then correct redshift $\tilde{z}_1$ is chosen such that physical evolution time is the same as the cosmological time between two redshifts as following
\begin{align}
	\Delta t = t_z~.
\end{align}
The ridshift $z_2$ can be determined by $z_2 = (1 + z_1) \eta - 1$ and $\Delta t$ can be solved from $\delta(\Delta t)$, which gives us the redshift-time calibration $(z_1, z_2, \Delta t)$ in the circular PBH binary systems.

In two parameter PBH binary systems, the standard timer can also be approached. We consider the PBH binaries from two different redshifts with their observed probability distribution $S_o(a_{z_1}, t_1)$ and $S_o(a_{z_2}, t_2)$. In the large major axis limit, the redshift ratio $\eta$ can be numerically solved in Eq.~\eqref{eq:redshift_ratio}. In the small major axis limit, we can numerically solve $\delta(\Delta t_z)$ as we have discussed in Sec.~\ref{sec:multi_parameter}. In order to obtain their redshifts $z_1$ and $z_2$, we also assume the cosmological time $t_z$ between two redshifts $z_1$ and $z_2$ follows the standard $\Lambda$CDM cosmology. Giving a redshift $\tilde{z}_1$, we can numerically solve $\Delta t$ from $\delta(\Delta t_z)$ and the right redshifts $z_1$ should satisfy $\Delta t = t_z~$. Then redshift-time calibration $(z_1, z_2, \Delta t)$ is constructed in two parameter PBH binary systems.

In future GW detections, after collecting a number of GWs from PBH binaries, we first classify them into different redshifts by their redshifted chirp mass. Choose the PBH binary systems from two different redshifts, their redshifts can be determined by comparing their redshifted probability distribution as we have discussed above, we can set one of obtained redshifts as a standard redshift $z_0$, the PBH mass can be recovered from the observed redshifted mass by $M_\mathrm{PBH} = M_z/(1 + z_0)$ and redshift of other probability distributions can be obtained by $z_\mathrm{PBH} = (1 + z_0) \eta - 1$. Then we can numerically solve physical evolution time $\Delta t$ between $z_0$ and $z_\mathrm{PBH}$ as discussed in Sec.~\ref{sec:multi_parameter}. Consequently, the redshift-time calibration $(z_0, z_\mathrm{PBH}, \Delta t)$ is obtained and the standard timer can be well developed without initial probability distribution.

Furthermore, cosmological models can be tested in standard timers. Considering the cosmological redshift-time relation $dz/dt = -(1 + z) H(z)$, we apply the obtained redshift-time calibration $(z_0, z_\mathrm{PBH}, \Delta t)$ from standard timers, which gives
\begin{align}
	\int_{z_0}^{z_\mathrm{PBH}} \frac{dz}{(1 + z) H(z)} = \int_{t_\mathrm{PBH}}^{t_0} dt = \Delta t~.
\end{align}
Take the flat $\Lambda$CDM model as an example, $H(z) = H_0 \sqrt{\Omega_\gamma(1 + z)^4 + \Omega_m(1 + z)^3 + \Omega_\Lambda}$. After constructing the redshift-time calibration in PBH binary systems from the primordial Universe to the present Universe, the Markov chain Monte Carlo (MCMC) simulation can be applied on the flat $\Lambda$CDM model in constraining the Hubble parameter $H_0$, energy density fraction of radiation $\Omega_\gamma$, matter $\Omega_m$, and cosmological constant $\Omega_\Lambda$.

\section{Conclusion and Discussions}\label{sec:conclusion_discussions}

To summarize, we propose that PBH binary systems can lead to standard timers to record the evolution of the cosmological redshift $z(t)$. Under the assumption of random distribution of PBHs in space, PBH binary systems have an identical initial probability distribution on major axis and eccentricity. By studying the evolution of the probability distribution in binary systems, the physical evolution time between the initial and later probability distribution can be extracted. Then the redshift-time relation can be constructed by studying the probability distribution of PBH binary systems at different redshifts. In order to obtain the probability distribution on major axis and eccentricity from the same redshift shell. We assume that PBH mass is monochromatic, through GWs produced from PBH binaries, their redshifted chirp mass can be obtained in GW waveforms, the PBH binaries from the same redshift have the same redshifted chirp mass and the mass ratio follows $q = 1$, then we extract the redshifted probability distribution on major axis and eccentricity from the same redshift.

For demonstrating how standard timers work in PBH binary systems, we perform an analytically study with a toy model in single parameter PBH binary systems where eccentricity is set as $e = 0$ and a numerically studied practical model in non-circular PBH binary systems. We show that the redshift can be determined by comparing the initial and redshifted probability distribution at the large major axis limit and the physical evolution time can be obtained by comparing the initial and recovered intrinsic probability distribution at the small major axis limit. Considering the initial probability distribution on major axis and eccentricity in PBH binary systems is indeterminate, the redshift of observed probability distributions cannot be directly obtained. We assume that the cosmological time between two redshifts follows the standard cosmology, then proper redshift of PBH binary systems should be chosen when the physical evolution time between two redshifted probability distributions equals its cosmological time, which further leads to standard timers in PBH binary systems without initial conditions.

In the above discussions, we mainly focus on PBH binary systems with a monochromatic mass spectrum, which helps classify the redshift of PBH binaries. In a general description of PBH binary systems, an extended mass spectrum should be taken into consideration, which is shown at the end of Sec.~\ref{sec:multi_parameter}. The standard timer can be constructed by a numerical study in a probability distribution on major axis, eccentricity and mass of PBHs in binaries $dP/da de dM_\mathrm{G}$. However, an extended mass spectrum of PBHs could cause difficulties in redshift classification of PBH binaries, which needs further studies in redshift identification, e.g., \cite{Namikawa:2015prh, Oguri:2016dgk, Osato:2018mtm} in standard sirens.

Also, standardization of PBH binaries as standard timers requires a detailed study on the initial conditions of PBH binaries, including primordial distributions on PBH mass and spin \cite{Mirbabayi:2019uph, DeLuca:2019buf, Green:2020jor}, initial spatial distribution of PBHs \cite{Tada:2015noa, Young:2015kda, Desjacques:2018wuu, Suyama:2019cst}, initial probability distribution on PBH binary parameters \cite{Sasaki:2016jop, Ioka:1998nz}, etc. Due to lack of observations on PBHs, initial conditions of PBH binaries are indeterminate, which would introduce the systematic uncertainty in calibration of redshift-time relation. Such a difficulty would be possibly overcome after future theoretical studies on PBH physics and observations on PBH signals in electromagnetic and GW channels. Then, PBH binary systems can work as standard timers in tracking cosmic evolution.

In general, the cosmological standard timer can be constructed based on dynamical systems in the Universe. Due to the same formation mechanism of dynamical systems, the statistical distribution of their initial states can be set as the standard reference, through the evolution mechanism of their statistical distribution, the elapsed time in the standard timer is evaluated. Meanwhile, cosmological redshift is encoded in the observable from dynamical systems. For signals from individual sources locally, the redshifted statistical distribution in dynamical systems from the same redshift can be obtained, and further gives their redshift by comparing with the initial state (see \cite{Cai:2021zxo,Cai:2021fgm} for more details). For signals from sources globally distributed in the Universe, as we have discussed in this article, GWs from PBH binaries globally, the redshifted statistical distribution from the same redshift can be extracted according to their redshifted parameters, and hence obtain their redshift. Consequently, the redshift-time calibration is constructed in a general dynamical system (see Appendix.~\ref{appendix:single_parameter} \& \ref{appendix:multi_parameter} for the detailed formalism).

\section*{Acknowledgements}

I thank Yi Wang for useful suggestions and discussions, Yi-Fu Cai, Chao Chen for useful discussions. I would like to thank Dr.~Xingwei Tang for her inspiration and encouragement.

\appendix

\section{The standard timer from single parameter dynamical systems}\label{appendix:single_parameter}

In constructing standard timers in dynamical systems, we need to set a particular condition of dynamical systems as a standard reference.  For example, the standard reference in standard candles is consistent peak luminosity produced by Type Ia supernovae (SNe), and the standard reference in standard rulers is the fixed baryon acoustic oscillation (BAO) scale that the sound wave can travel before the recombination. Generally, initial states of dynamical systems are uncertain under the Gaussian distribution of perturbations. However,  the statistical distribution of initial states in dynamical systems can be unique due to the same physical mechanism behind them, which can be set as a standard reference. With the standard reference, the physical evolution time and redshift can be extracted by studying the evolution of observed dynamical systems, which can help calibrate $z(t)$.

For simplicity, we start with a single parameter dynamical system, whose time evolution follows $dM/dt = -f(M)$. Here $M$ is the observable physical parameter that characterizes the dynamical system and $f(M)$ is its time derivative function. The statistical distribution of the single parameter dynamical system $S(M; t)$ can be described as
\begin{align}\label{eq:distribution}
	S(M; t) = \frac{dN}{dM_t}~,
\end{align}
where $M_t \equiv M(t)$, $N$ is the statistic of the distribution of dynamical systems. In order to trace the historical evolution of dynamical systems, a standard initial distribution is essential. Eq.~\eqref{eq:distribution} can be written as
\begin{align}\label{eq:time_dist}
	S(M; t) = \frac{dN}{dM_\mathrm{i}}\frac{dM_\mathrm{i}}{dM_t}~.
\end{align}
Here, $dN/dM_\mathrm{i}$ is the initial statistical distribution of dynamical systems $S(M; t_\mathrm{i})$. $dM_\mathrm{i}/dM_t$ describes the evolution of the dynamical system, which can be further expressed by its time evolution $dM/dt = -f(M)$, which gives
\begin{align}\label{eq:time_evolution}
	\int_{M_\mathrm{i}}^{M_t}\frac{dM}{f(M)} = g(M_t) - g(M_\mathrm{i}) = -\Delta t~.
\end{align}
Here, function $g(M)$ is an antiderivative of function $1/f(M)$. Then, the evolution of dynamical systems can be written as
\begin{align}\label{eq:sys_evolution}
	\frac{dM_\mathrm{i}}{dM_t} = \frac{g'(M_t)}{g'(M_\mathrm{i})} = \frac{g'(M_t)}{g'(g^{-1}(g(M_t) + \Delta t))}~.
\end{align}
Here, $g'(M) \equiv dg(M)/dM$ and $g^{-1}$ denotes the inverse function of $g(M)$. As the result, Eq.~\eqref{eq:time_dist} can be further expressed as
\begin{align}\label{eq:time_dist_detail}
	S(M; t) = \frac{dN}{dM_\mathrm{i}}\frac{g'(M_t)}{g'(g^{-1}(g(M_t) + \Delta t))}~.
\end{align}
The physical evolution time $\Delta t$ can be extracted by giving an initial statistical distribution $dN/dM_\mathrm{i}$ in Eq.~\eqref{eq:time_dist_detail}.

In the observational aspect, the statistical distribution of the dynamical system is deformed, due to observed physical parameter $M$ is redshifted by the cosmological expansion, which gives the observational distribution $S_o(M_z; t)$ (subscript $o$ denotes the observational quantity) as
\begin{align}\label{eq:observed_single_paremeter}\nonumber
	S_o(M_z; t) &= \frac{dN}{dM_\mathrm{i}(z)}\frac{dM_\mathrm{i}(z)}{dM_z}\\ 
	&=  \frac{dN}{dM_\mathrm{i}(z)}\frac{g'(M_z)}{g'(g^{-1}(g(M_z) + \Delta t_z))}~.
\end{align}
Here, $M_z$ denotes the redshifted physical parameter, such as redshifted photon energy $E_z = E/(1+z)$ and redshifted chirp mass in binary black hole systems $\mathcal{M}_z = (1+z)\mathcal{M}$. $dN/dM_\mathrm{i}(z)$ characterizes the cosmological redshift effect in the initial statistical distribution. Following Eq.~\eqref{eq:time_evolution}, we have $g(M_\mathrm{i}(z)) = g(M_z) + \Delta t_z$.

In order to extract the redshift-time calibration, we consider two cases in Eq.~\eqref{eq:observed_single_paremeter}. For a fixed evolution time, the first case is $g(M_z) \gg \Delta t_z$, which makes sure the time evolution is negligible and redshift can be extracted by comparing the redshifted physical parameter $M_z$ with the initial physical parameter $M_\mathrm{i}$ in the initial statistical distribution. The second case is $g(M_z) \ll \Delta t_z$, where $\Delta t_z$ dominates in the redshifted physical parameter, which gives $g(M_\mathrm{i}(z)) \simeq \Delta t_z$. Then $\Delta t_z$ can be extracted in following expression
\begin{align}\label{eq:redshift-time}\nonumber
	&S_o(M_z; t) \simeq \\
	&\left\{
	\begin{aligned}
		&\frac{dN}{dM_\mathrm{i}(z)} \qquad \qquad \quad \qquad \quad~,~g(M_z) \gg \Delta t_z \\
		&\frac{dN}{dg^{-1}(\Delta t_z)} \frac{g'(M_z)}{g'(g^{-1}(\Delta t_z))} \quad ,~g(M_z) \ll \Delta t_z 
	\end{aligned}
	\right.
\end{align}

Above all, we have discussed the formalism of a standard timer in an observable dynamical system $S(M; t)$. However, this formalism does not apply to the case that $M$ is not an observable of dynamical systems, meanwhile, the signals produced from them is an observable, e.g., electromagnetic waves and gravitational waves. In these cases, we consider the following integral equation,
\begin{align}
	P(E; t) = \int_0^{\infty}K(E, M) S(M; t)dM~.
\end{align}
Here, $K(E, M)$ is the kernel function which transfers an unobservable distribution $S(M; t)$ to an observable distribution $P(E; t)$. $S(M; t)$ can be extracted by an inverse integral equation
\begin{align}
	S(M; t) = \int_0^{\infty}K^{-1}(E, M) P(E; t)dE~,
\end{align}
where $K^{-1}(E, M)$ is the inverse kernel function of $K(E, M)$. Due to the cosmological expansion, the observed physical parameter
$E$ is redshifted to $E_z$. Therefore, the observable becomes
\begin{align}\label{eq:obs_dist}
	P_o(E_z; t) = \int_0^{\infty}K_o(\mathcal{Z}_1(E_z), M)S(M; t)dM~,
\end{align}
where $\mathcal{Z}_1$ function describes the redshift effect in the observable $E_z$. In order to construct the redshift-time relation, a redshift term need to appear in $S(M; t)$, which requires the connection between $E$ and $M$ in the kernel function, for instance, the primary Hawking radiation kernel follows $H(E(1+z), M) = H(E, M(1+z))$ \cite{Cai:2021fgm}. Therefore, we assume the kernel function follows
\begin{align}\label{eq:kernel_transfer}
	K_o(\mathcal{Z}_1(E_z), M) = K_o(E_z, \mathcal{Z}_2(M))~.
\end{align}
Here, $\mathcal{Z}_1$ and $\mathcal{Z}_2$ function describe how the redshift term transfers from $E_z$ to $M$ in kernel function. Then, Eq.~\eqref{eq:obs_dist} can be written as
\begin{align}
	P_o(E_z; t) = \int_0^{\infty}K_o(E_z, \mathcal{Z}_2(M))S_o(\mathcal{Z}_2(M); t) d\mathcal{Z}_2(M)~.
\end{align}
As the result, $S_o(\mathcal{Z}_2(M); t)$ is given by
\begin{align}\label{eq:inverse_trans}
	S_o(\mathcal{Z}_2(M); t) = \int_0^{\infty}K_o^{-1}(E_z, \mathcal{Z}_2(M)) P_o(E_z; t) dE_z~.
\end{align}
As we discuss in Eq.~\eqref{eq:redshift-time}, $S_o(\mathcal{Z}_2(M); t)$ can be expressed in two conditions in Eq.~\eqref{eq:two_cond_z}, which further gives the redshift-time calibration.
\begin{align}\label{eq:two_cond_z}\nonumber
	&S_o(\mathcal{Z}_2(M); t) \simeq \\
	&\left\{
	\begin{aligned}
		&\frac{dN}{d\mathcal{Z}_2(M_\mathrm{i})} \qquad \qquad \qquad \qquad ,~g(\mathcal{Z}_2(M)) \gg \Delta t_z \\
		&\frac{dN}{dg^{-1}(\Delta t_z)} \frac{g'(\mathcal{Z}_2(M))}{g'(g^{-1}(\Delta t_z))} \quad ~,~g(\mathcal{Z}_2(M)) \ll \Delta t_z
	\end{aligned}
	\right. 
\end{align}

\section{The standard timer from multi-parameter dynamical systems}\label{appendix:multi_parameter}

In general, we consider multi-parameter dynamical systems in the Universe, whose statistical distribution can be expressed as
\begin{align}\label{eq:multi-para}
	S(\mathbf{M}; t) = \frac{dN}{d^n\mathbf{M}_t}~,
\end{align}
where $\textbf{M}$ denotes the $n$-dimensional physical parameter vector which characterizes the dynamical system. By introducing the initial statistical distribution as the standard reference, Eq.~\eqref{eq:multi-para} can be written as
\begin{align}\label{eq:multi_time_dist}
	S(\mathbf{M}; t) = \frac{dN}{d^n\mathbf{M}_\mathrm{i}} \det{\mathbf{J}(\mathbf{M},\Delta t)}~.
\end{align}
Here, $\mathbf{J}$ is the Jacobian of the dynamical system which is defined as $\mathbf{J}_{ij} \equiv \partial M_i(t_\mathrm{i})/\partial M_j(t)$.  With the reference of the initial statistical distribution, the physical evolution time $\Delta t$ can be extracted from the determinant of the Jacobian $\det{\mathbf{J}(\mathbf{M}, \Delta t)}$. However, due to strong coupling among different parameter components in its time evolution $d\mathbf{M}/dt = -\mathbf{f}(\mathbf{M})$, the general analytical expression of the Jacobian element $\partial M_i(t_\mathrm{i})/\partial M_j(t)$ can hardly be found, which indicates the numerical solution of $\det{\mathbf{J}(\mathbf{M}, \Delta t)}$ is essential in extracting the physical evolution time $\Delta t$.

In the observational perspective, the redshift term caused by the cosmological expansion also appears in the statistical distribution of multi-parameter dynamical systems as following
\begin{align}
	S_o(\mathbf{M}_z; t) = \frac{dN}{d^n\mathbf{M}_\mathrm{i}(z)} \det{\mathbf{J}(\mathbf{M}_z, \Delta t_z)}~,
\end{align}
where $\mathbf{M}_z$ denotes the redshifted $n$-dimensional physical parameter vector and $dN/d^n\mathbf{M}_\mathrm{i}(z)$ characterizes the redshifted initial statistical distribution.

As we have discussed in Appendix.~\ref{appendix:single_parameter}, we consider two cases in extracting the redshift-time calibration. One case is that in the parameter space where the time evolution of parameters is negligible compared with their initial value, which gives $\det{\mathbf{J}(\mathbf{M}_z, \Delta t_z)} \simeq 1$. Then the redshift $z$ can be obtained by comparing the observed statistical distribution $S_o(\mathbf{M}_z; t) \simeq dN/d^n\mathbf{M}_\mathrm{i}(z)$ with the initial one. The other case is that in the parameter space where the time evolution of parameters dominates their initial value, where physical evolution time $\Delta t$ can be extracted from the numerical solution.

In the scenario that $S(\mathbf{M}; t)$ is not observable, we consider the observable $P(E; t)$ as following
\begin{align}
	P(E; t) = \int_{V}K(E, \mathbf{M}) S(\mathbf{M}; t) d^n \mathbf{M}~,
\end{align}
where, $K(E, \mathbf{M})$ is the transfer kernel which transfers an unobservable $S(\mathbf{M}; t)$ to an observable $P(E; t)$, $V$ is the integral region of $n$-dimensional parameter $\mathbf{M}$. With the expansion of the Universe, the redshift effect appears in the observable in the following form
\begin{align}
	P_o(E_z; t) = \int_{V} K_o(\mathcal{Z}_1(E_z), \mathbf{M}) S(\mathbf{M}; t) d^n \mathbf{M}~.
\end{align}
As we have shown in Eq.~\eqref{eq:kernel_transfer}, we introduce $\mathcal{Z}_1$ and $\mathcal{Z}_2$ function to transfer a redshift term from $E_z$ to $\mathbf{M}$, which gives
\begin{align}
	P_o(E_z; t) = \int_{V} K_o(E_z, \mathcal{Z}_2(\mathbf{M})) S_o(\mathcal{Z}_2(\mathbf{M}); t) d^n \mathcal{Z}_2(\mathbf{M})~.
\end{align}
As the result, the unobservable statistical distribution $S_o(\mathcal{Z}_2(\mathbf{M}); t)$ can be obtained by an inverse kernel transformation as Eq.~\eqref{eq:inverse_trans},
\begin{align}
	S_o(\mathcal{Z}_2(\mathbf{M}); t) = \int_{0}^{\infty} K_o(E_z, \mathcal{Z}_2(\mathbf{M}))^{-1} P_o(E_z; t) d E_z~.
\end{align}
However, the analytical form of the inverse kernel in multi-parameter dynamical systems $K_o(E, \mathbf{M})^{-1}$ could hardly be found, which needs further numerical methods, e.g., the method for the least squares problem \cite{lawson1995solving, PROVENCHER1982213}. After obtaining $S_o(\mathcal{Z}_2(\mathbf{M}); t)$, the redshift-time calibration can be extracted as Eq.~\eqref{eq:two_cond_z}.

\end{document}